\documentclass{article}
\usepackage{graphicx}
\usepackage{amssymb}
\begin{document}
\title{\textbf{Quantum Stephani Universe in vicinity of the symmetry center}}
\author{Pouria Pedram\thanks{pouria.pedram@gmail.com}\\{\small Department of Physics,
Shahid Beheshti University, Evin, Tehran 19839, Iran}\\
 {\small Research Center for Quantum Computing,
Interdisciplinary Graduate School of
Science and Engineering, }\\{\small Kinki University, Higashi-Osaka, Osaka 577-8502, Japan}}
\date{\today}
\maketitle \baselineskip 24pt
\begin{abstract}
We study a class of spherically symmetric Stephani cosmological
models in the presence of a self-interacting scalar field in both
classical and quantum domains. We discuss the construction of
`canonical' wave packets resulting from the solutions of a class of
Wheeler-DeWitt equations in the Stephani Universe. We suggest
appropriate initial conditions which result in wave packets
containing some desirable properties, most importantly good
classical and quantum correspondence. We also study the situation
from de-Broglie Bohm interpretation of quantum mechanics to recover
the notion of time and compare the classical and Bohmian results. We
exhibit that the usage of the canonical prescription and appropriate
choices of expansion coefficients result in the suppression of the
quantum potential and coincidence between classical and Bohmian
results. We show that, in some cases, contrary to
Friedmann-Robertson-Walker case, the bound state solutions also
exist for all positive values of the cosmological constant.
\end{abstract}
\textit{Pacs}:{ 98.80.Qc, 04.40.Nr, 04.60.Ds;}

\section{Introduction}\label{sec1}
In recent years observations show that the expansion of the Universe
is accelerating  in the present epoch \cite{1} contrary to
Friedmann-Robertson-Walker (FRW) cosmological models, with
non-relativistic matter and radiation. Some different physical
scenarios using exotic form of matter have been suggested to resolve
this problem \cite{2,3,4,5,6,pedramIJTP}. In fact the presence of
exotic matter is not necessary to drive an accelerated expansion.
Instead we can relax the assumption of the homogeneity of space,
leaving the isotropy with respect to one point. The most general
class of non-static, perfect fluid solutions of Einstein's equations
that are conformally flat is known as the ``Stephani Universe''
\cite{1-1,2-1}. This model can be embedded in a five-dimensional
flat pseudo-Euclidean space \cite{1-1,9,3-3} and its three
dimensional spatial sections are homogeneous and isotropic
\cite{12}. Recently, quantum spherically symmetric Stephani
cosmological models in the presence of the perfect fluid have been
studied in Refs. \cite{pedramPLB,pedramCQG2}. In these works the
Schutz's variational formalism \cite{Schutz1,Schutz2} is applied to
recover the notion of time and investigate the singularity avoidance
at the quantum level.

The question of construction and interpretation of wave packets in
quantum cosmology and its connection with classical cosmology has
been attracting much attention in recent years. Moreover, numerous
studies have been done to obtain a quantum theory for gravity and to
understand its connection with classical physics.

In quantum cosmology, in analogy with ordinary quantum mechanics,
one is generally concerned with the construction of wave functions
by the superposition of the `energy eigenstates' which would peak
around the classical trajectories in configuration space, whenever
such classical-quantum correspondence is mandated by the nature of
the problem. However, contrary to ordinary quantum mechanics, a
parameter describing time is absent in quantum cosmology. Therefore,
the initial conditions would have to be expressed in terms of an
{\it intrinsic} time parameter, which in the case of the
Wheeler-DeWitt (WDW) equation could be taken as the local scale
factor for the three geometry \cite{dewitt}. Also, since the sign of
the kinetic term for the scale factor is negative, a formulation of
the Cauchy problem for the WDW equation is possible. The existence
of such a sign is one of the exclusive features of gravity with many
other interesting implications.

The construction of wave packets resulting from the solutions of the
WDW equation has been a common feature of some research works in
quantum cosmology
\cite{kiefer,tucker,sepangi,wavepacket,pedramCQG1}. In particular,
in references \cite{wavepacket,pedramCQG1,pedramPLB3} the
construction of wave packets in a Friedmann Universe is presented in
detail and appropriate boundary conditions are motivated. Generally
speaking, one of the aims of these investigations has been to find
wave packets whose probability distributions coincide with the
classical paths obtained in classical cosmology. In these works, the
authors usually consider theories in which a self interacting scalar
field is coupled to gravity in a Robertson-Walker type Universe. The
solutions are obtained such that the following desirable properties
are satisfied. There should be a good classical-quantum
correspondence, which means that the wave packet should centered
around the classical path, the crest of the wave packet should
follow as closely as possible the classical path, and to each
distinct classical path there should correspond a wave packet with
the above properties. Recently, a general prescription has been
suggested by Gousheh \textit{et al} \cite{pedramCQG1} for
constructing the `canonical' wave packets which contain all above
desired properties. They showed that there always exists a
``canonical initial slope" (CIS) for a given initial wave function,
which optimizes some desirable properties of the resulting wave
packet, most importantly good classical-quantum correspondence.

In this paper we deal with the subject of ``initial condition''
which is an important problem in quantum cosmology. In fact, in
classical cosmology we can uniquely determine the classical initial
conditions subject to the zero energy condition. But in quantum
cosmology since the underlying equation (WDW equation) is a
hyperbolic differential equation, we are free to choose the initial
wave function and the initial derivative of the wave function by
choosing arbitrary expansion coefficients. These quantities
(distributions) correspond to classical initial position and initial
momentum, respectively. Therefore, although WDW equation allows us
to use different choices of initial conditions upon choosing
different expansion coefficients, these wave functions correspond to
different classical situations. This also happens whenever WDW-like
equation appears in other theories such as varying speed of light
quantum cosmological models \cite{pedramPLB3}. Hence, a legitimate
question which arises is how we can construct a specific wave packet
which completely corresponds to its unique classical counterpart?
One possible solution is removing the arbitrariness of the expansion
coefficients and defining a certain relation between them. In our
previous investigations \cite{pedramCQG1,pedramPLB3} we discovered
that given a particular choice of initial wave function, certain
coefficients remain undetermined, and if we set the functional form
of those coefficient to be the same as the determined ones, we
obtain excellent classical and quantum correspondence.

Here, we are interested to study the Stephani Universe in the
presence of a scalar field. First, we write the reduced action near
symmetry center ($r\approx0$) and find the corresponding
hamiltonian. Then we obtain the Einstein's equations and WDW
equations in minisuperspace. These equations can be solved
numerically with appropriate initial conditions. In particular, we
use Spectral Method (SM) \cite{pedramCPC} as an accurate and stable
numerical method for solving the quantum cosmology case which can be
cast in the form of a hyperbolic PDE. The form of the scalar
potential is chosen to contain some desirable properties like the
cosmological constant, positive mass term in Taylor expansion, and
to be bounded from below. Then, we construct the wave packets for
various functional forms of the spatial curvature through the
canonical prescription \cite{pedramCQG1}.

The paper is organized as follows: In Sec.~\ref{sec2}, we outline
the main problem which is a case of Stephani cosmology where the
matter is taken to be a particular type of self-interacting scalar
field. We derive the main equations both for the classical cosmology
and the quantum cosmology. We begin Sec.~\ref{sec3} with a
description of the Spectral Method \cite{pedramCPC} which is a
robust numerical method, and then we review the general prescription
of canonical wave packets. We then consider various cases and solve
them in both classical and quantum cosmological domains. In
Sec.~\ref{sec3b}, we find the corresponding Bohmian trajectories and
compare the classical and quantum solutions. In Sec.~\ref{sec4}, we
draw some final conclusions.

\section{The model}\label{sec2}
Let us start from the Einstein-Hilbert action plus a scalar field as
\begin{eqnarray}
\label{action} S =\frac{1}{2}\int_Md^4x\sqrt{-g}\, {\cal R} +
2\int_{\partial M}d^3x\sqrt{h}\, h_{ab}\, K^{ab}+
\int_Md^4x\sqrt{-g}\, \left(-\frac{1}{2}(\nabla
\phi)^2-U(\phi)\right),
\end{eqnarray}
where $K^{ab}$ is the extrinsic curvature and $h_{ab}$ is the
induced metric over the three-dimensional spatial hypersurface,
which is the boundary $\partial M$ of the four dimensional manifold
$M$ in units where $8\pi G=1$ \cite{7-2}. The last term of
(\ref{action}) represents the scalar field contribution to the total
action.

The metric in spherically symmetric Stephani Universe
\cite{1-1,2-1,12,9,10,14} has the following form
\begin{eqnarray}\label{metric}
ds^2 = -N^2(r,t)dt^2+
\frac{R^2(t)}{V^2(r,t)}\left[dr^2+r^2(d\theta^2+\sin^2\theta
\,d\phi^2)\right],
\end{eqnarray}
where
\begin{eqnarray}\label{lapse}
N(r,t)=F(t) \frac{\displaystyle R(t)}{\displaystyle
V(r,t)}\frac{\displaystyle\partial}{\displaystyle\partial
t}\left(\frac{\displaystyle V(r,t)}{\displaystyle R(t)}\right),
\end{eqnarray}
is the lapse function and the functions $V(r,t)$ and $F(t)$ are
defined as
\begin{eqnarray}
  V(r,t) &=& 1+\frac{1}{4}k(t)r^2, \\
  F(t) &=& \frac{R(t)}{\sqrt{C^2(t)R^2(t)-k(t)}},
\end{eqnarray}
where $k$, $R$, and $C$ are arbitrary functions of time
\cite{Dabrowski,Stelmach}. Here, $k(t)$ plays the role of the
spatial curvature and $R(t)$ is the Stephani version of the FRW
scale factor. Although $k(t)$ is an arbitrary function of time in
the Stephani model, assuming a power law relation between $R(t)$ and
$k(t)$ makes the model solvable and is in agreement with the
accelerating expansion of the Universe
\cite{pedramPLB,pedramCQG2,14,Stelmach,Godlowski,Sussman}. However,
some authors have used some thermodynamics relations to obtain a
power law relation between these two variables \cite{Sussman}.
Though in the spherical symmetric inhomogeneous models, the scalar field
$\phi$ depends on both $r$ and $t$, we can consider it as an only
function of time near the symmetry center $r\approx0$ which means
$\phi=\phi(t)$ \cite{inhomogeneous1,inhomogeneous2,inhomogeneous3}.

By substituting the Stephani metric (\ref{metric}) in the action
(\ref{action}) and choosing the curvature function $k(t)$ in the
form \cite{pedramPLB,pedramCQG2,14,Stelmach,Godlowski,Sussman}
\begin{equation}\label{k(t)}
k(t)=\beta R^{\gamma}(t),
\end{equation}
after dropping the surface terms and with due attention to the form
of the lapse function (\ref{lapse}), the final reduced action near
$r\approx0$ takes the form
\begin{equation}
S = \int dt\left[-3\frac{\displaystyle\dot{R}^2R}{\displaystyle N}
+3\beta\,N\,
R^{1+\gamma}+N\,R^3\left(\frac{1}{2}\dot\phi^2-U(\phi)\right)
\right].
\end{equation}
Now choosing the gauge $N=1$ \cite{Dabrowski,Stelmach}, we have the
following Lagrangian
\begin{equation}\label{lagrangian}
L=-3\dot{R}^2R +3\beta
R^{1+\gamma}+R^3\left(\frac{1}{2}\dot\phi^2-U(\phi)\right).
\end{equation}
Therefore, in this limit, the Stephani Universe is equivalent to the
FRW model where the curvature term can be chosen as an arbitrary
function of time.

The Einstein's equations for $r\approx0$ resulting from above
Lagrangian with the zero energy condition can be written as
\begin{eqnarray}
3\left[\left(\frac{\dot{R}}{R}\right)^2+\beta R^{\gamma-2}\right]&=&
\frac{\dot{\phi}^2}{2}+U(\phi), \label{eq4}\\
2\left(\frac{\ddot{R}}{R}\right)+\left(\frac{\dot{R}}{R}\right)^2+
\beta (1+\gamma)R^{\gamma-2}&=&-\frac{\dot{\phi}^2}{2}+U(\phi), \label{eq5}\\
\ddot{\phi}+3\frac{\dot{R}}{R}\dot{\phi}+\frac{\partial
U}{\partial\phi} &=&0, \label{eq6}
\end{eqnarray}
where dot represents differentiation with respect to time. We
require the potential $U(\phi )$ to have natural characteristics for
small $\phi$, so that we may identify the coefficient of
$\frac{1}{2}\phi^2$ in its Taylor expansion as a positive mass
squared $m^2$, and $U(0)$ as a cosmological constant $\Lambda$. An
interesting choice of $U(\phi)$ with three free parameters is
\cite{tucker,wavepacket,pedramCQG1,darabi1,darabi2,ahmadi}
\begin{eqnarray}
U(\phi)=\Lambda+\frac{m^2}{2\alpha^2}\sinh^2(\alpha\phi)+
\frac{b}{2\alpha^2}\sinh(2\alpha\phi).  \label{eq7}
\end{eqnarray}
In the above expression $m^2=\partial^2 U/\partial\phi^2|_{\phi=0}$
is a mass squared parameter and $b$ is a coupling constant. We need
to choose $\alpha^2=\frac{\displaystyle3}{\displaystyle8}$ in order
to separate the variables in the Lagrangian. This potential is
bounded from below and as we shall see, prevents us from usual
problem of factor ordering. Moreover, since this type of potential
also has been used for FRW cosmological models, we can compare our
solutions with the previous FRW results
\cite{wavepacket,pedramCQG1}.

The Lagrangian (\ref{lagrangian}) can be cast into a simple form
using the transformations $X=R^{3/2} \cosh(\alpha\phi)$ and
$Y=R^{3/2} \sinh(\alpha\phi)$, which transform the term $R^3 U(\phi
)$ into a quadratic form. Upon using a second transformation to
eliminate the coupling term in the quadratic form, we arrive at new
variables $u$ and $v$, which are linear combinations of $X$ and $Y$
\begin{eqnarray}
\left(
\begin{array}{l}
 u \\
 v
\end{array}
\right)=\left(
\begin{array}{rr}
 \cosh\theta & \sinh\theta \\
 \sinh\theta & \cosh\theta
\end{array}
\right)\left(
\begin{array}{l}
 X \\
 Y
\end{array}
\right),
\end{eqnarray}
where
\begin{eqnarray}
\theta=\frac{1}{2} \tanh ^{-1}\left(\frac{2 b}{m^2}\right).
\end{eqnarray}
In terms of the new variables, the Lagrangian takes on the following
simple form
\begin{eqnarray}
L(u,v)=\frac{4}{3}\left[ (\dot{u}^2-\omega_1^2
u^2)-(\dot{v}^2-\omega_2^2 v^2)-\frac{9}{4}\beta
(u^2-v^2)^{(\gamma+1)/3} \right],\label{eq9}
\end{eqnarray}
where $\omega_{1,2}^2 = -3\Lambda /4 +m^2/2 \mp\sqrt{m^4-4b^2}/2$.
The resulting Einstein's equations are
\begin{eqnarray}
\ddot{u}+\omega_1^2u+\frac{3\beta}{4}(\gamma+1)u(u^2-v^2)^{(\gamma-2)/3}=0,\label{u''}
\end{eqnarray}
\begin{eqnarray}
\ddot{v}+\omega_2^2v+\frac{3\beta}{4}(\gamma+1)v(u^2-v^2)^{(\gamma-2)/3}=0,\label{v''}
\end{eqnarray}
\begin{eqnarray}
\dot{u}^2+\omega_1^2u^2-\dot{v}^2-\omega_2^2v^2+\frac{9}{4}\beta
(u^2-v^2)^{(\gamma+1)/3} =0.\label{u2}
\end{eqnarray}
Equations (\ref{u''}) and (\ref{v''}) are the dynamical equations
and (\ref{u2}) is the zero energy constraint. The non-linearity of
these equations for $\gamma\ne2$ is now apparent. The corresponding
quantum cosmology is described by the Wheeler-DeWitt equation
written as
\begin{eqnarray}
H\psi(u,v)=\left\{-\frac{\partial^2}{\partial u^2}+\frac{\partial^2}
{\partial v^2}+ \omega_1^2 u^2-\omega_2^2 v^2+\frac{9}{4}\beta
(u^2-v^2)^{(\gamma+1)/3}\right\}\psi(u,v)=0, \label{eq10}
\end{eqnarray}
which arises from the zero energy condition (\ref{u2}). In general,
this equation  is not exactly solvable and we should resort to a
numerical method \cite{pedramCPC}.

\section{Solutions for the quantum cosmology cases}\label{sec3}
We start this section by a discussion of the numerical method that
we shall use and then we outline the general prescription for
finding CIS. The general hyperbolic PDE that we want to solve is
$$\left\{-\frac{\partial^2}{\partial u^2}+\frac{\partial^2}
{\partial v^2}+ \hat{f}'(u,v)\right\}\psi(u,v)=0,$$ where
$\hat{f}'(u,v)$ is an arbitrary function. It is notable that such
equations may represent a wave-like equation whose solution may
rapidly oscillate. In such cases, the usual spatial integration
routines such as Finite Difference Methods fail to produce a
reasonable solution. Therefore, it is of prime importance to use a
reliable, efficient and accurate numerical method \cite{pedramCPC}.

SM \cite{SP1} consists of first choosing a complete orthonormal set
of eigenstates of a preferably relevant hermitian operator to
construct the solution. Since the whole set of the complete basis
has usually infinite elements, we make the approximation of
representing the solution by only a finite superposition of the
basis functions. By substituting this approximate  solution into the
differential equation, a matrix equation is obtained. The expansion
coefficients of these approximate solutions could be determined by
the eigenfunctions of this matrix. In this method, the accuracy of
the solution is increased by choosing a larger set of basis
functions. Having resorted to a numerical method, it is worth
setting up a more general problem defined by the following WDW
equation
\begin{eqnarray}
H\psi(u,v)=\left\{-\frac{\partial^2}{\partial u^2}+\frac{\partial^2}
{\partial v^2}+ \omega_1^2 u^2-\omega_2^2
v^2+\hat{f}(u,v)\right\}\psi(u,v)=0. \label{eq10new}
\end{eqnarray}
As mentioned before, any complete orthonormal set can be used. Here
we use the Fourier series basis by restricting the configuration
space to a finite square region of sides $2L$. This means that we
can expand the solution as
\begin{eqnarray}
\psi(u,v)=\sum_{i,j=1}^2 \sum_{m,n} A_{m,n,i,j} \,\,\,
g_i\left(\frac{m \pi u}{L}\right)\,\,\, g_j\left(\frac{n \pi
v}{L}\right), \label{eqpsitrigonometric}
\end{eqnarray}
where
\begin{equation}\label{eqsincos}
\left\{
  \begin{array}{ll}
    g_1\left(\frac{m \pi
u}{L}\right)=\sqrt{\frac{2}{R_{m} L}}\sin\left(\frac{m \pi
u}{L}\right), &  \\
   g_2\left(\frac{m \pi
u}{L}\right)=\sqrt{\frac{2}{R_{m} L}}\cos\left(\frac{m \pi
u}{L}\right). & \\
  \end{array}
\right. \mbox{and}\,\,\, R_{m} =\left\{ \begin{array}{ll}
    1, & m\ne0 \\
    2, & m=0 \\
\end{array}
\right.
\end{equation}
By referring to the WDW equation (\ref{eq10new}), we realize that in
the Fourier basis it is appropriate to introduce $\hat f'$ as
\begin{eqnarray}
\hat f'(u,v)=\hat{f}(u,v)+\omega_1^2 u^2-\omega_2^2 v^2.\label{eqf'}
\end{eqnarray}
We can make the following expansion
\begin{eqnarray}\label{eqf'2}
\hat f'(u,v) \psi(u,v)=\sum_{i,j} \sum_{m,n} B'_{m,n,i,j} \,\,\,
g_i\left(\frac{m \pi u}{L}\right)\,\,\, g_j\left(\frac{n \pi
v}{L}\right),
\end{eqnarray}
where $B'_{m,n,i,j} $ are coefficients that can be determined once
$\hat f'(u,v)$ is specified. By substituting
(\ref{eqpsitrigonometric},\ref{eqf'2}) in (\ref{eq10new}) and using
the independence of $g_i\left(\frac{m \pi u}{L}\right)$s and
$g_j\left(\frac{n \pi v}{L}\right)$s we obtain
\begin{eqnarray}
\left[\left(\frac{m \pi }{L}\right)^2 -\left(\frac{n \pi
}{L}\right)^2\right] A_{m,n,i,j}+B'_{m,n,i,j}=0,\label{eqAB'}
\end{eqnarray}
where
\begin{eqnarray}
B'_{m,n,i,j}\hspace{-3mm}&=& \sum_{m',n',i',j'}
\left[\int_{-L}^{L}\int_{-L}^{L} g_{i}\left(\frac{m \pi u}{L}\right)
g_{j}\left(\frac{n \pi v}{L}\right) \hat f'(u,v)
g_{i'}\left(\frac{m' \pi u}{L}\right) g_{j'}\left(\frac{n' \pi
v}{L}\right)du dv\right]A_{m',n',i',j'}\nonumber\\ &=&
\sum_{m',n',i',j'} C'_{m,n,i,j,m',n',i',j'}\,\,
A_{m',n',i',j'}.\label{B'}
\end{eqnarray}
Therefore we can rewrite (\ref{eqAB'}) as
\begin{eqnarray}
\left[\left(\frac{m \pi}{L} \right)^2 -\left(\frac{n \pi}{L}\right
)^2\right] A_{m,n,i,j}+ \sum_{m',n',i',j'}
C'_{m,n,i,j,m',n',i',j'}\,\, A_{m',n',i',j'}=0.\label{eqAC'}
\end{eqnarray}
Now, we select $4N^2$ basis functions, that is $m$ and $n$ run from
$1$ to $N$. It is obvious that the presence of the operator $\hat
f'(u,v)$ leads to nonzero coefficients $C'_{m,n,i,j,m',n',i',j'}$ in
(\ref{eqAC'}), which in principle could couple all of the matrix
elements of $A$. Then we replace the square matrix $A$ with a column
vector $A'$ with $(2N)^2$ elements, so that any element of $A$
corresponds to one element of $A'$. This transforms (\ref{eqAC'}) to
\begin{eqnarray}
D\, A'=0. \label{eqmatrix2}
\end{eqnarray}
Matrix $D$ is a square matrix with $(2N)^2 \times (2N)^2$ elements
which can be obtained from (\ref{eqAC'}). Equation (\ref{eqmatrix2})
can be looked as an eigenvalue equation, {\it i.e.} $DA'_a=a A'_a$
with $(2N)^2$ eigenvectors. However, for constructing the acceptable
wave functions, {\em i.e. }the ones satisfying the WDW equation
(\ref{eq10new}), we only require eigenvectors which span the null
space of the matrix $D$. That is, due to (\ref{eqAC'}) we will have
exactly $2N$ null eigenvectors which will be linear combination of
our original eigenfunctions introduced in
(\ref{eqpsitrigonometric}). After finding the $2N$ eigenvectors of
$D$ with zero eigenvalue, {\em i.e.} $A'^k$ ($ k=1,2,3,...,2N$), we
can find the corresponding elements of matrix $A$, $A^k_{m,n,i,j}$.
Therefore, the wave function can be expanded as
\begin{eqnarray}
\psi(u,v)= \sum_{k} \lambda^k \psi^k(u,v)=\sum_{k} \lambda^k
\sum_{m,n,i,j} A^k_{m,n,i,j}\,\,\, g_i\left(\frac{m \pi
u}{L}\right)\,\,\, g_j\left(\frac{n \pi v}{L}\right).
\label{eqpsifinal2}
\end{eqnarray}
where $\lambda^k\,\,$s are arbitrary complex coefficients which can
be fixed by the initial conditions.

We are free to adjust two parameters: $2N$, the number of basis
elements and $2L$, the length of the spatial region. This length
should be preferably larger than spatial spreading of all the sought
after wave functions. However, if $2L$ is chosen to be too large we
loose overall accuracy. Therefore, it is important to note that for
each $N$, $L$ should be properly adjusted \cite{SP1}.

Now to determine $\lambda^k\,\,$s we need to apply the initial
conditions. As a mathematical point of view, since the underling
differential equation is second order, $\lambda^k\,\,$s are
arbitrary and independent coefficients. On the other hand, if we are
interested in constructing the wave packets which simulate the
classical behavior with known classical positions and velocities,
these coefficients will not be all independent yet. To address this
issue, let us study the problem near the solution's boundary
($v=0$). We can approximate (\ref{eq10}) near the $v=0$, so up to
the first order in $v$ we have
\begin{eqnarray}
\left\{-\frac{\partial^2}{\partial u^2}+\frac{\partial^2} {\partial
v^2}+ \omega_1^2 u^2+\frac{9}{4}\,\beta\,
u^{2(\gamma+1)/3}\right\}\psi(u,v)=0. \label{eq10nearv0}
\end{eqnarray}
This PDE is separable in $u$ and $v$ variables, so we can write
\begin{equation}\label{psi-separated}
\psi(u,v)=\varphi^{\gamma}(u)\chi(v).
\end{equation}
By substituting $\psi(u,v)$ in (\ref{eq10nearv0}), two ODEs can be
derived
\begin{eqnarray}
\frac{d^2\chi_n(v)}{dv^2}+E_n\chi_n(v)&=&0,
\label{eqseparated1}\\
\hspace{-0.6cm}-\frac{d^2\varphi_n^{\gamma}(u)}{d
u^2}+\left(\omega_1^2 u^2+\frac{9}{4}\,\beta\,
u^{2(\gamma+1)/3}\right)\varphi_n^{\gamma}(u)&=&E_n\varphi_n^{\gamma}(u),\label{eqseparated2}
\end{eqnarray}
where $E_n$s are separation constants. These equations are
Schr\"{o}dinger-like equations with $E_n$s as their `energy' levels.
Equation (\ref{eqseparated1}) is exactly solvable with plane wave
solution as
\begin{equation}\label{eqplanewave}
\chi_n(v)=\alpha_n\cos\left(\sqrt{E_n}\,\,v\right)+i\beta_n\sin\left(\sqrt{E_n}\,\,v\right),
\end{equation}
where $\alpha_n$ and $\beta_n$ are arbitrary complex numbers.
Equation (\ref{eqseparated2}) does not seem to be exactly solvable
and we resort to a numerical technique. As mentioned before, SM can
be used to find the bound state energy levels ($E_n$) and the
corresponding wave functions ($\varphi_n(u)$) with high accuracy.
The general solution to the (\ref{eq10nearv0}) can be written as
\begin{equation}\label{psi-separated2}
\psi(u,v)=\sum_{n=\mbox{\footnotesize{even}}} (A_n
\cos(\sqrt{E_n}v)+iB_n\sin(\sqrt{E_n}v))
\varphi_n^{\gamma}(u)+\sum_{n=\mbox{\footnotesize{odd}}}( C_n
\cos(\sqrt{E_n}v)+iD_n\sin(\sqrt{E_n}v)) \varphi_n^{\gamma}(u).
\end{equation}
The separation of this solution to even and odd terms, though in
principle unnecessary, is crucial for our prescription for the CIS.
As stated before, this solution is valid only for small $v$. It is
obvious that the presence of the odd terms of $v$ dose not have any
effect on the form of the initial wave function but they are
responsible for the slope of the wave function at $v=0$, and vice
versa for the even terms. The general initial conditions can now be
written as
\begin{eqnarray}
\psi(u,0)&=&\sum_{even}A_n\varphi_n^{\gamma}(u)+\sum_{odd}C_n\varphi_n^{\gamma}(u)\label{eqinitial1}\\
\psi'(u,0)&=&i\sum_{even}B_n\sqrt{E_n}\,\,\varphi_n^{\gamma}(u)+i\sum_{odd}D_n\sqrt{E_n}\varphi_n^{\gamma}(u),\label{eqinitial2}
\end{eqnarray}
where prime denotes the derivative with respect to $v$. Obviously a
complete description of the problem would include the specification
of both these quantities. However, given only the initial condition
on the wave function, we show there is a CIS which produces a
canonical wave packet with all the aforementioned desired
properties. We can qualitatively describe the prescription for this
case as setting the functional form of the odd undetermined
coefficients to be the same as the even determined ones and vice
versa. This means that the coefficients that determine CIS i.e.
$B_n$ for $n$ even and $D_n$ for $n$ odd, are chosen as
\cite{pedramCQG1}
\begin{equation}\label{eqcanonicalslope}
  B_n=C_n\,\,\, \,\,\,\,\mbox{for $n$ even}\hspace{1cm}D_n=A_n\,\,\,\,\,\, \mbox{for $n$
  odd}
\end{equation}
In other word, by specifying the initial wave function, the
prescription (\ref{eqcanonicalslope}) automatically construct the
appropriate initial slope which coincide well with the classical
counterpart. Note that, although $C_n$ ($A_n$) is defined only for
$n$ odd (even), we can extend its definition to $n$ even (odd) by
choosing the same functional form. Now, using the canonical initial
conditions (\ref{eqinitial1},\ref{eqinitial2}), we can determine
$\lambda$s and construct the wave packet via equation
(\ref{eqpsifinal2}).

The classical paths corresponding to these solutions can be obtained
from (\ref{u''},\ref{v''}). The corresponding initial conditions for
the classical case are
\begin{eqnarray}
 u(0)=u_0,\hspace{0.5cm} v(0)=0,\hspace{0.5cm}
\dot{u}(0)=0,\hspace{0.5cm}
\dot{v}(0)=\dot{v}_0,\label{eqclassicalic}
\end{eqnarray}
where the parameters $u_0$ and $\dot{v}_0$ are adjusted so that the
zero energy condition (\ref{u2}) is satisfied.

For ease of comparison with FRW models \cite{pedramCQG1}, we choose
the same illustrative problem with $\omega_1=\omega_2\equiv \omega$.
Moreover, for all studied cases in this section ($\gamma=2,4,6$) we
choose the same coefficients as
\begin{equation}
  A_n=e^{-\frac{1}{4} |\chi|^2}\frac{\chi^n}{\sqrt{2^n n!}}\,\,\, \,\,\,\,\mbox{for $n$ even}\hspace{1cm}B_n=0\,\,\,\,\,\, \mbox{for $n$
  odd}\label{coef}
\end{equation}
where $\chi$ is a free parameter. This choice of expansion
coefficients obviously result in different initial conditions for
various values of $\gamma$ (\ref{eqinitial1}). Note that, $\gamma=0$
case is equivalent to FRW case with the constant spatial curvature
\cite{pedramCQG1}.

\begin{figure}
\centerline{\begin{tabular}{ccc}
\includegraphics[width=8cm]{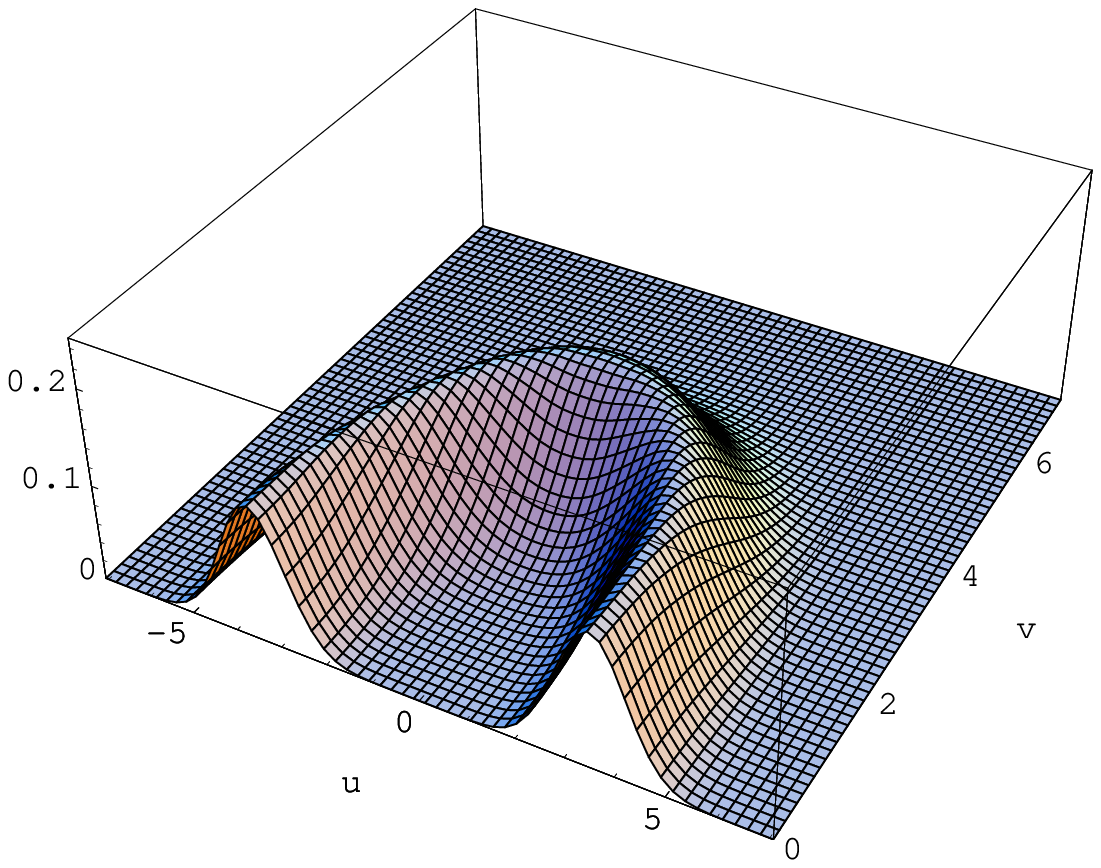}
 &\hspace{2.cm}&
\includegraphics[width=6.5cm]{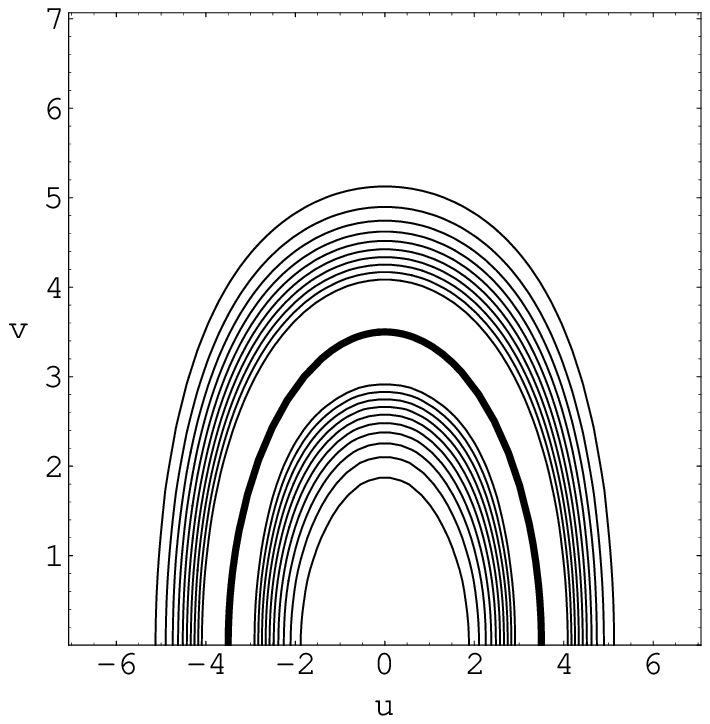}
\end{tabular}}
\caption{$\gamma=2$ case: Left, the square of the wave packet $|
\psi(u,v)|^2$ for $\omega^2=1$, $\beta=1$, $\chi=3.5$ and $N=15$.
Right, the contour plot of the same figure with the classical path
superimposed as the thick solid line.} \label{fig1}
\end{figure}

For $\gamma=2$, the Lagrangian (\ref{lagrangian}) can be written as
\begin{eqnarray}
L=-3\dot{R}^2R
+R^3\left(\frac{1}{2}\dot\phi^2-(U(\phi)-3\beta)\right),
\end{eqnarray}
which is equivalent to the flat FRW cosmological model, but with a
modified cosmological constant, $\Lambda'=\Lambda-3\beta$. For this
case the WDW equation (\ref{eq10}) reduces to
\begin{eqnarray}
\left\{-\frac{\partial^2}{\partial u^2}+\frac{\partial^2} {\partial
v^2}+ \omega_1^2 u^2-\omega_2^2 v^2\right\}\psi(u,v)=0,
\end{eqnarray}
This equation is in the form of an isotropic
oscillator-ghost-oscillator and is separable in the configuration
space variables. The general solution can thus be written as a sum
over the product of simple harmonic oscillator wave functions with
the same frequencies. The exact classical paths would be Lissajous
figures in general. In particular, for $\omega_1=\omega_2$, by using
the expansion coefficients in the form of equation (\ref{coef}), the
corresponding classical paths are circles with radii $\chi$. The
result is shown in the left part of Fig.~\ref{fig1}. As can be seen
from this figure, the parameters of the problem are chosen such that
the initial state consists of two well separated peaks and this
class of problems are the ones which are also amenable to a
classical description. We should mention that there are a variety of
different cases illustrated in Ref.~\cite{wavepacket} including
$\omega_1\ne \omega_2$. Having precisely set the initial conditions
for both the classical and quantum cosmology cases, we can now
superimpose the results as illustrated in the right part of the
Fig.~\ref{fig1}. As can be seen from the figure, the
classical-quantum correspondence is manifest.

\begin{figure}
\centerline{\begin{tabular}{ccc}
\includegraphics[width=8cm]{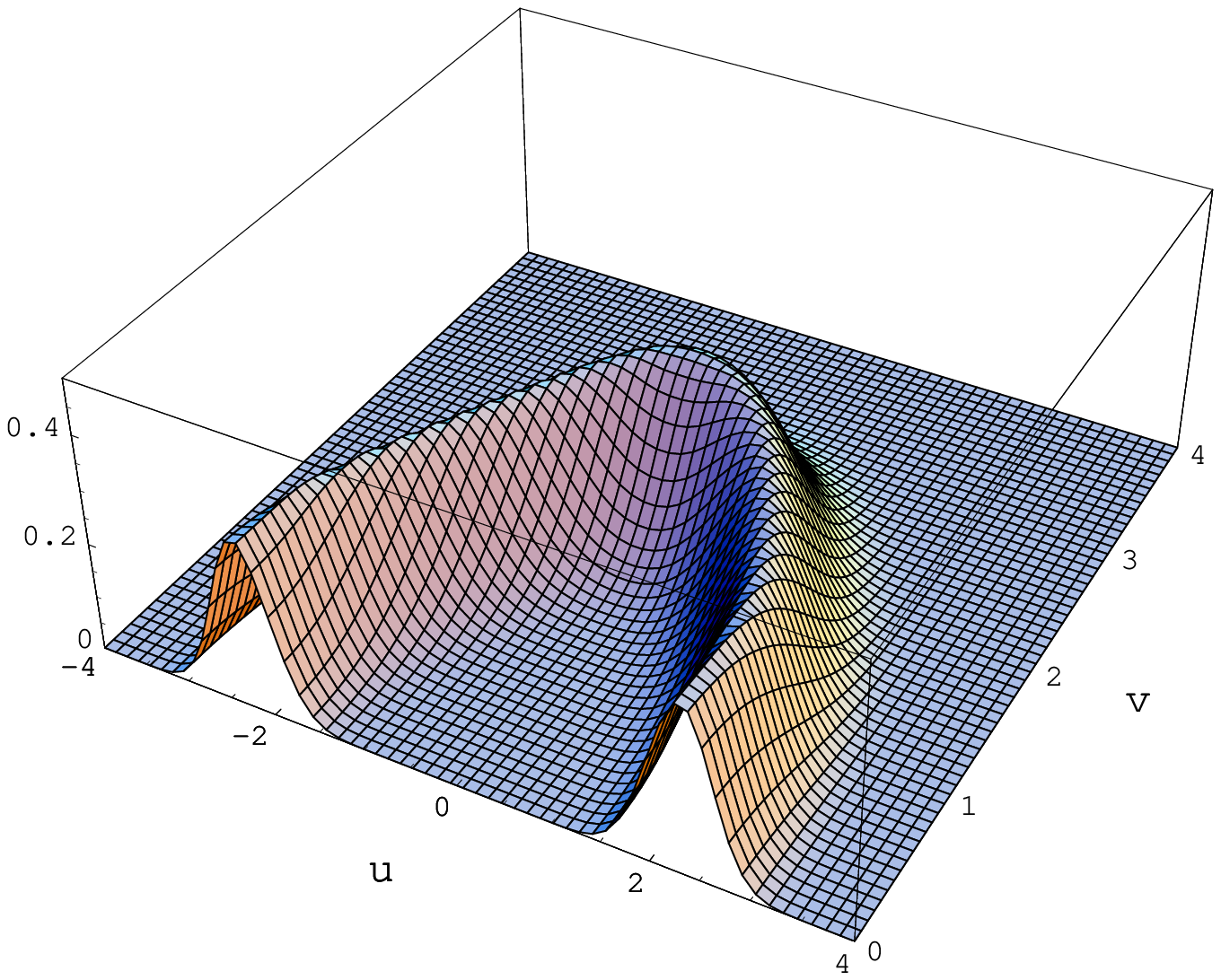}
 &\hspace{2.cm}&
\includegraphics[width=6.5cm]{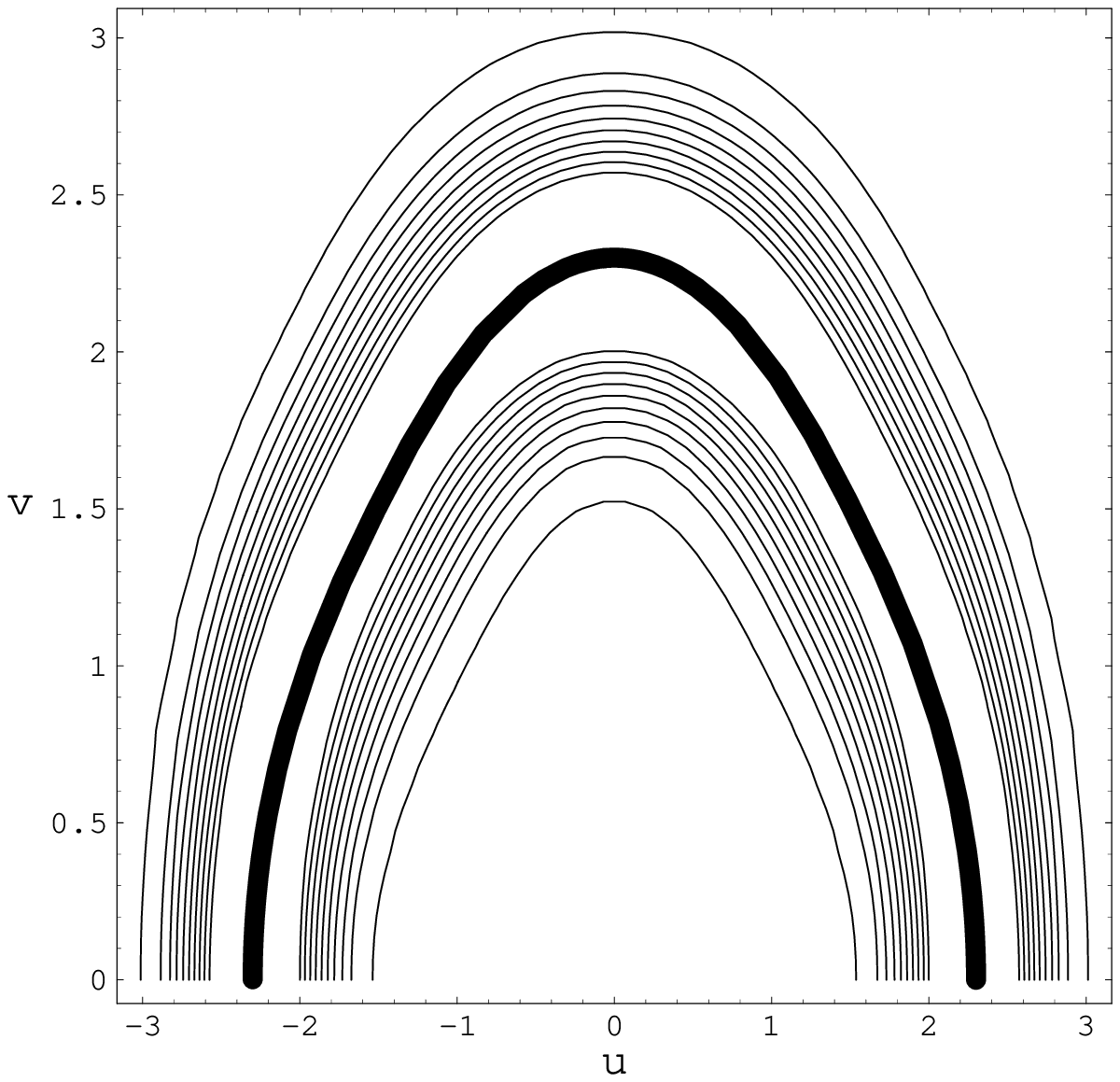}
\end{tabular}}
\caption{$\gamma=4$ case: Left, the square of the wave packet $|
\psi(u,v)|^2$ for $\omega^2=1$, $\beta=1$, $\chi=4$ and $N=15$.
Right, the contour plot of the same figure with the classical path
superimposed as the thick solid line.} \label{fig2}
\end{figure}

\begin{figure}
\centerline{\begin{tabular}{ccc}
\includegraphics[width=8cm]{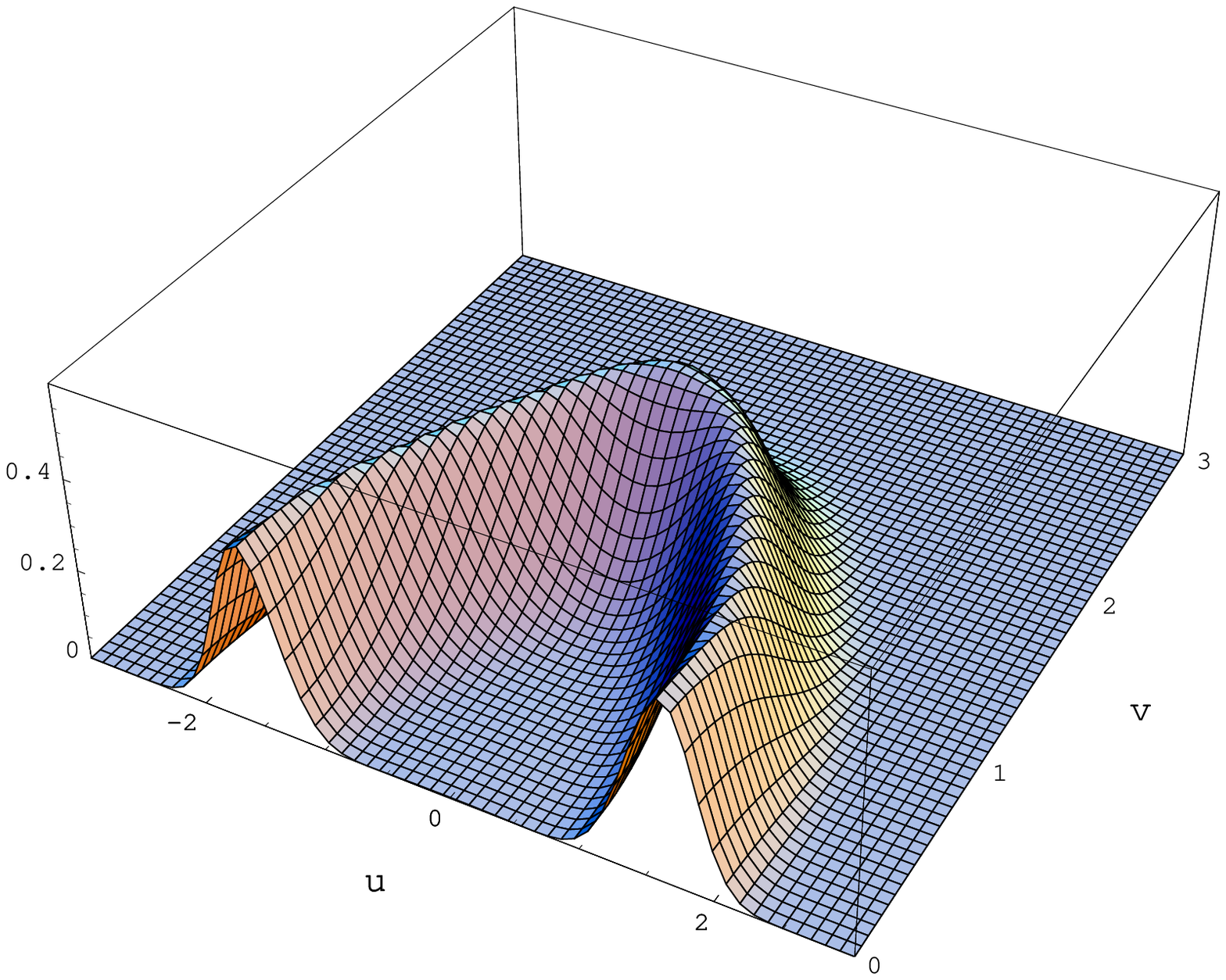}
 &\hspace{2.cm}&
\includegraphics[width=6.5cm]{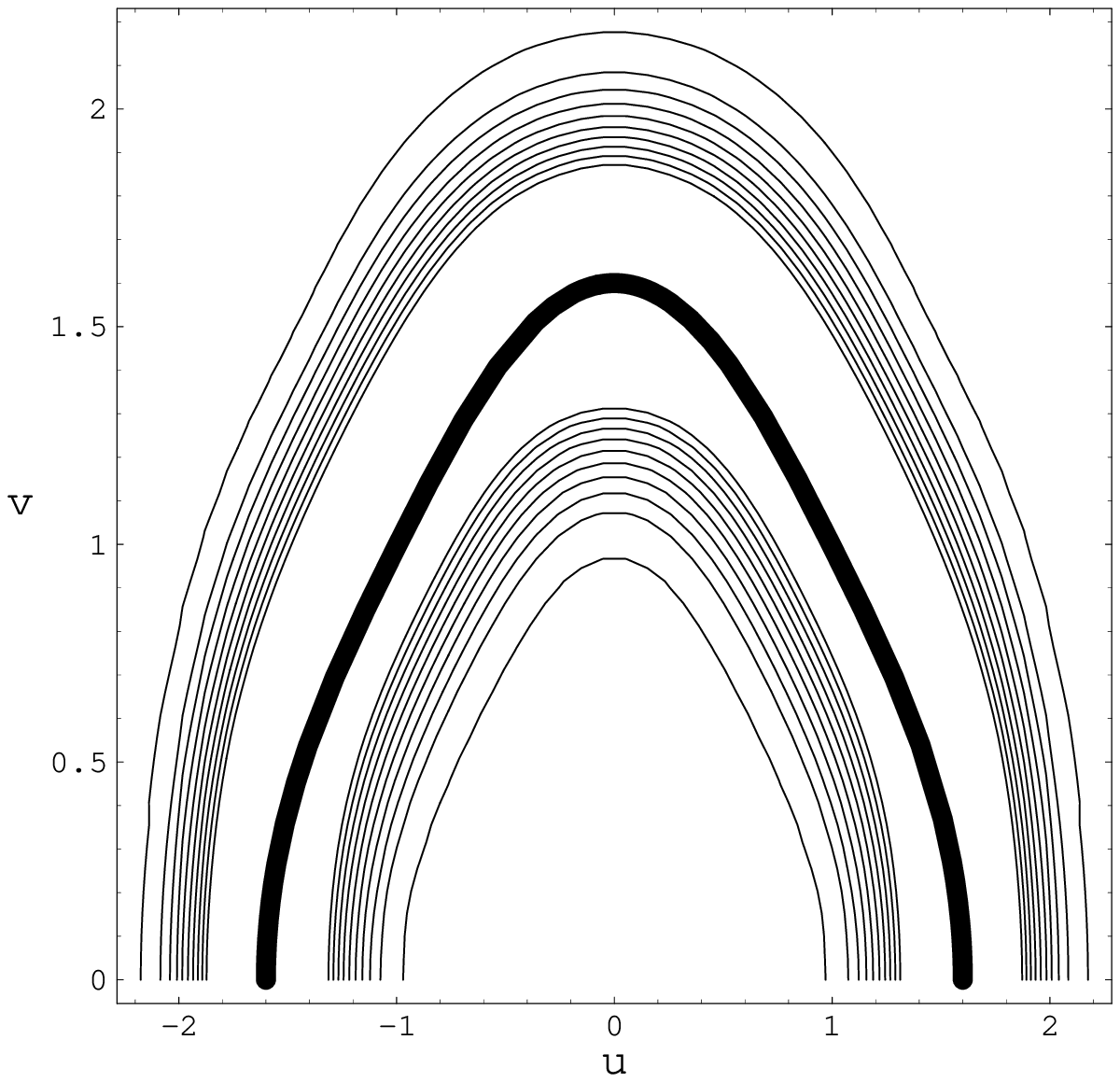}
\end{tabular}}
\caption{$\gamma=6$ case: Left, the square of the wave packet $|
\psi(u,v)|^2$ for $\omega^2=1$, $\beta=1$, $\chi=3$ and $N=15$.
Right, the contour plot of the same figure with the classical path
superimposed as the thick solid line.} \label{fig3}
\end{figure}

For $\gamma=4,6$, the WDW equation (\ref{eq10}) and the
corresponding classical field equations
(\ref{u''},\ref{v''},\ref{u2}) are not exactly solvable. The
classical equations can be solved numerically using customary
algorithms, and the quantum cases using SM. In fact, for $\gamma>2$,
the bound state solutions exist only for positive values of $\beta$
($\beta\geq0$). This is contrary to FRW case, where the bound state
solutions can be obtained for positive, zero, and negative values of
the spatial curvature \cite{pedramCQG1}. Now, using the canonical
prescription we can construct the wave packets which follow their
counterpart classical trajectories. Figures \ref{fig2},\ref{fig3}
show the resulting canonical wave packets and their classical
trajectories for $\gamma=(4,6)$, and $\chi=(3.5,4)$, respectively.
We have set $\beta=1$, and used $N=15$ basis functions to reproduce
the wave packets. Note that, for these cases the parameter $\chi$
corresponds to the classical initial position, but unlike the
previous case, the classical pathes are no longer circles.

\begin{figure}
\centerline{\begin{tabular}{ccc}
\includegraphics[width=8cm]{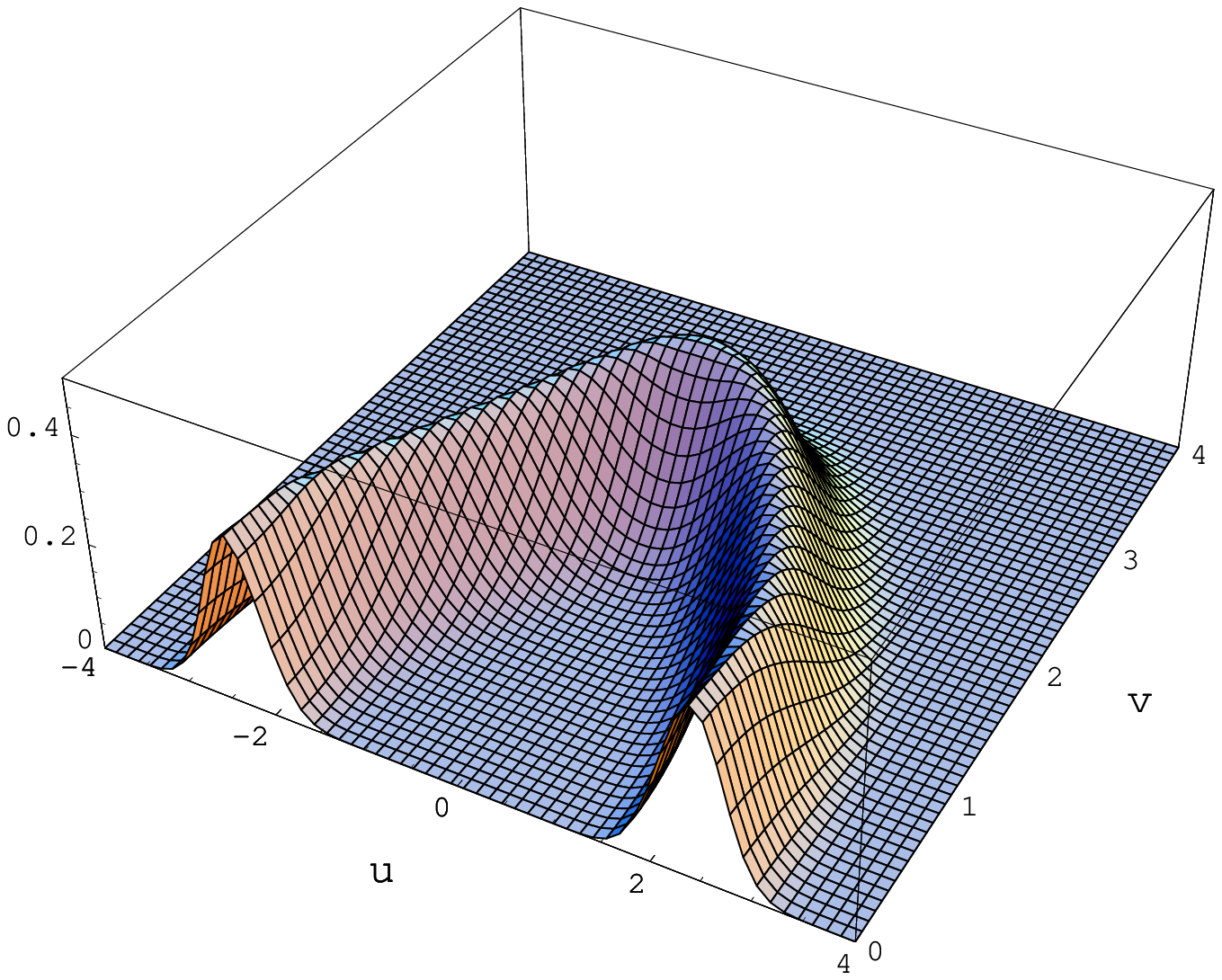}
 &\hspace{2.cm}&
\includegraphics[width=6.5cm]{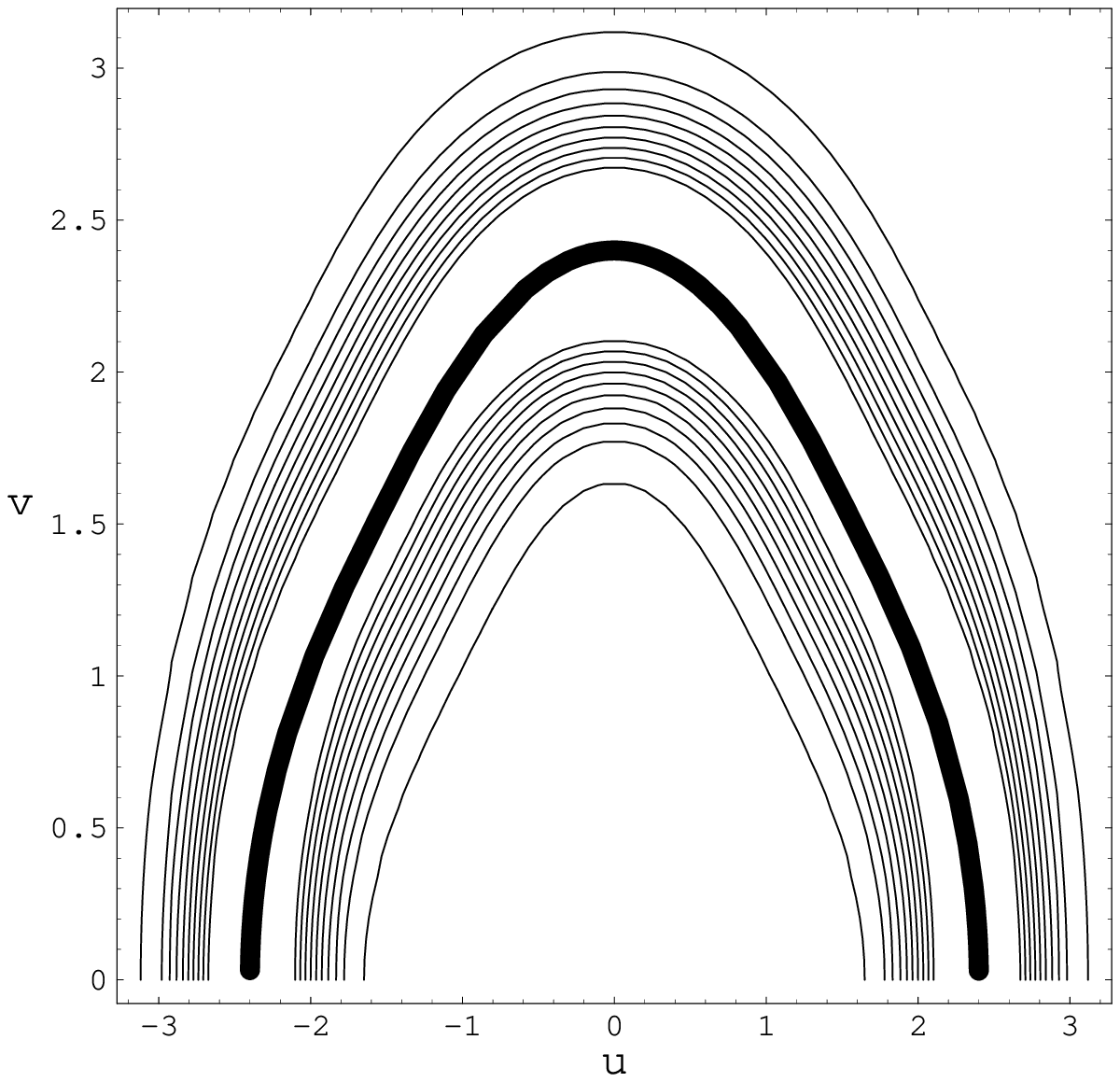}
\end{tabular}}
\caption{$\gamma=4$ case: Left, the square of the wave packet $|
\psi(u,v)|^2$ for $\omega^2=-1$, $\beta=1$, $\chi=4$ and $N=15$.
Right, the contour plot of the same figure with the classical path
superimposed as the thick solid line.} \label{fig4}
\end{figure}

An interesting feature of the Stephani Model is that it allows us to
still have bound state solutions even with negative values of
$\omega^2$. In fact, for $\gamma\geq2$ and $\beta>0$, bound state
solutions also exist for all positive values of $\Lambda$. Figure
\ref{fig4} shows the resulting classical and quantum mechanical
solutions for $\gamma=4$, $\omega^2=-1$, and $\beta=1$. Using the
same expansion coefficients (\ref{coef}), we have a slightly larger
initial position description with respect to the previous case where
$\gamma=4$, $\omega^2=1$, and $\beta=1$ (Fig.~\ref{fig2}). To be
more specific, the classical initial positions for these cases are
$u_0=2.3,2.4$ for $\omega^2=+1,-1$, respectively.

\section{Causal interpretation}\label{sec3b}
To make the connection between the classical and quantum results
more concrete, we can use de Broglie-Bohm interpretation of quantum
mechanics. In this interpretation the wave function is written as
\begin{equation}\label{R,S}
\Psi(u,v) = R\, e^{iS},
\end{equation}
where $R=R(u,v)$ and $S=S(u,v)$ are real functions and satisfy the
following equations
\begin{eqnarray}
\label{hje}  \left(\frac{\partial S}{\partial u}\right)^2+\omega_1^2 u^2
-\frac{1}{R}\frac{\partial^2 R}{\partial u^2}-
\left(\frac{\partial S}{\partial v}\right)^2-\omega_2^2 v^2+\frac{1}{R}\frac{\partial^2 R}{\partial v^2}+\frac{9}{4}\beta(u^2-v^2)^{(\gamma+1)/3} &=& 0,\\
R\frac{\partial^2 S}{\partial u^2} -R\frac{\partial^2 S}{\partial
v^2}+2 \frac{\partial R}{\partial u} \frac{\partial S}{\partial
u}-2\frac{\partial R}{\partial v} \frac{\partial S}{\partial v}&=&
0.
\end{eqnarray}
To write $R$ and $S$, it is more appropriate to separate the real
and imaginary parts of the wave packet
\begin{equation}
\Psi (u,v)=x(u,v)+i y(u,v),
\end{equation}
where $x,y$ are real functions of $u$ and $v$. Using (\ref{R,S}) we
have
\begin{eqnarray}
R&=&\sqrt{x^2+y^2},\\
S&=&\arctan(\frac{y}{x}).
\end{eqnarray}
On the other hand, the Bohmian trajectories are governed by
\begin{eqnarray}
p_u = \frac{\partial S}{\partial u},\\
p_v = \frac{\partial S}{\partial v},
\end{eqnarray}
where $p_u$ and $p_v$ are the momenta conjugate to $u$ and $v$
variables, respectively. Therefore, the Hamiltonian constraint
($H=0$) is again satisfied, but in the presence of the modified
potential (\ref{hje}). The Bohmian equations of motion take the form
\begin{eqnarray}
\dot{u}= \frac{1}{2}\frac{1}{1+\left(\frac{y}{x}\right)^2}\frac{d}{du}\left(\frac{y}{x}\right),\\
\dot{v}=-\frac{1}{2}\frac{1}{1+\left(\frac{y}{x}\right)^2}\frac{d}{dv}\left(\frac{y}{x}\right),
\end{eqnarray}
where $x$ and $y$ are known functions of $u$ and $v$
(\ref{psi-separated2}). These differential equations can be solved
numerically to find the time evolution of $u$ and $v$.

\begin{figure}
\centerline{\begin{tabular}{ccc}
\includegraphics[width=8cm]{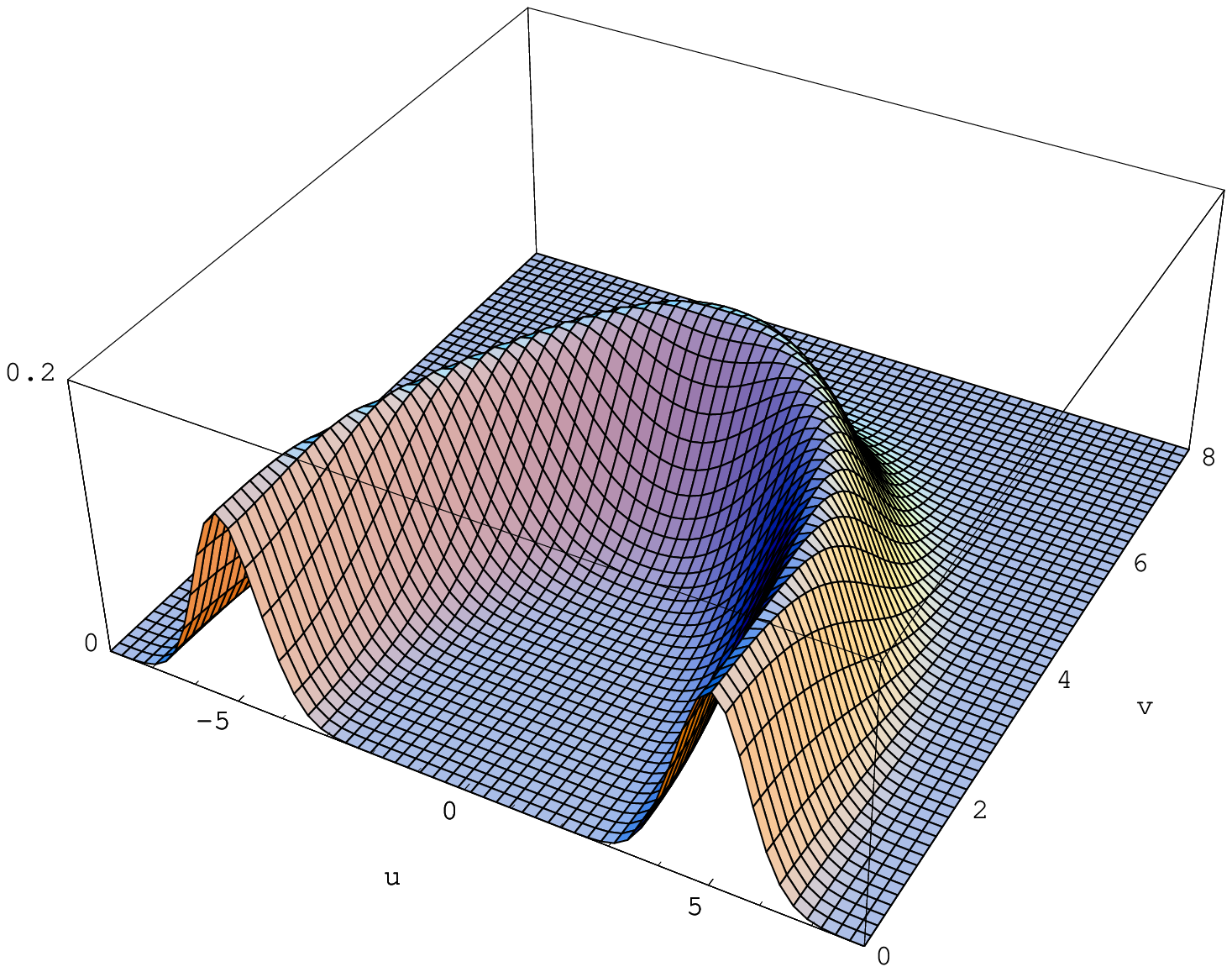}
 &\hspace{2.cm}&
\includegraphics[width=8cm]{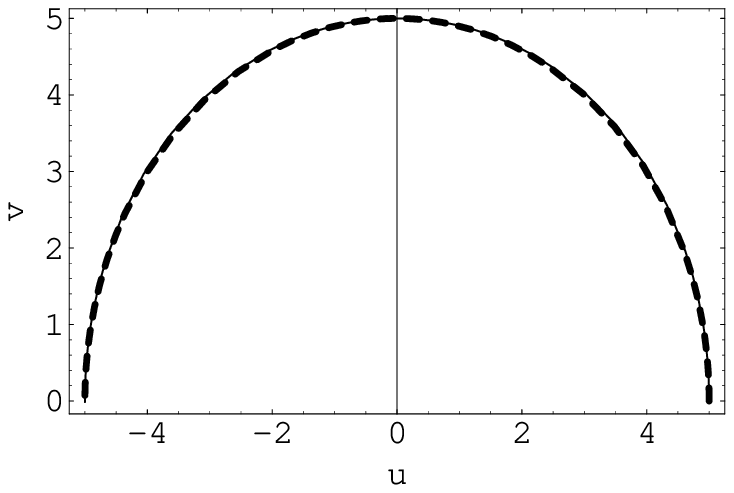}
\end{tabular}}
\caption{$\gamma=2$ case: Left, the square of the wave packet $|
\psi(u,v)|^2$ for
$A(n)=\frac{\,\chi^n}{\,{\sqrt{2^n\,n!}}}e^{-\chi^2/4}$ and
$\chi=5$. Right, the classical (dashed line) and Bohmian (solid
line) trajectories.} \label{fig1b}
\end{figure}

\begin{figure}
\centerline{\begin{tabular}{ccc}
\includegraphics[width=8cm]{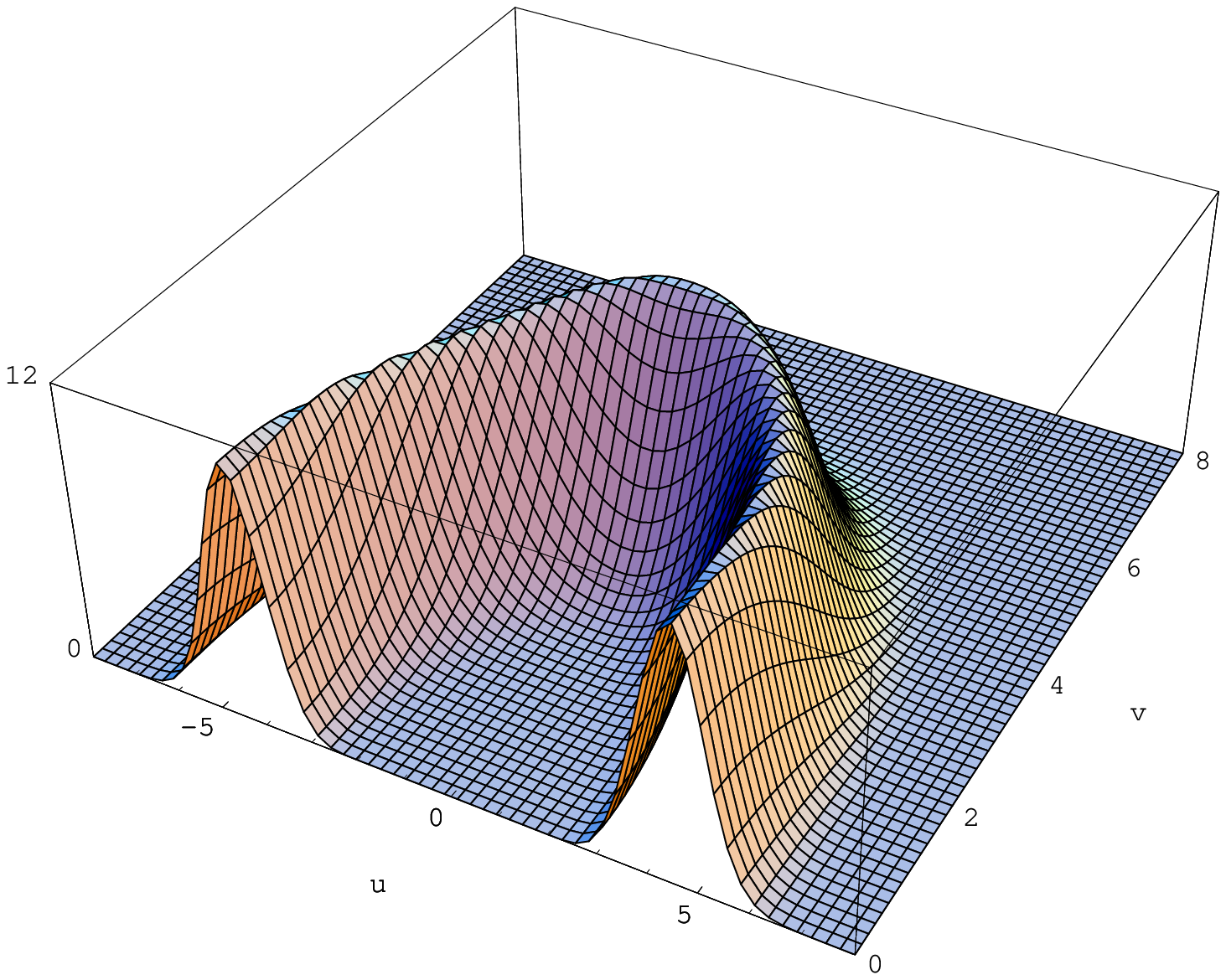}
 &\hspace{2.cm}&
\includegraphics[width=8cm]{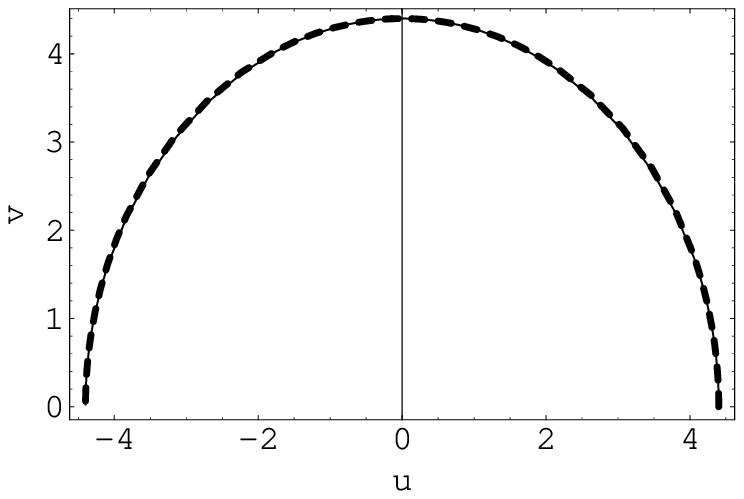}
\end{tabular}}
\caption{$\gamma=2$ case: Left, the square of the wave packet $|
\psi(u,v)|^2$ for
$A(n)=\frac{n\,\chi^n}{\,{\sqrt{2^n\,n!}}}e^{-\chi^2/4}$ and
$\chi=4$. Right, the classical (dashed line) and Bohmian (solid
line) trajectories.} \label{fig2b}
\end{figure}

\begin{figure}
\centerline{\begin{tabular}{ccc}
\includegraphics[width=8cm]{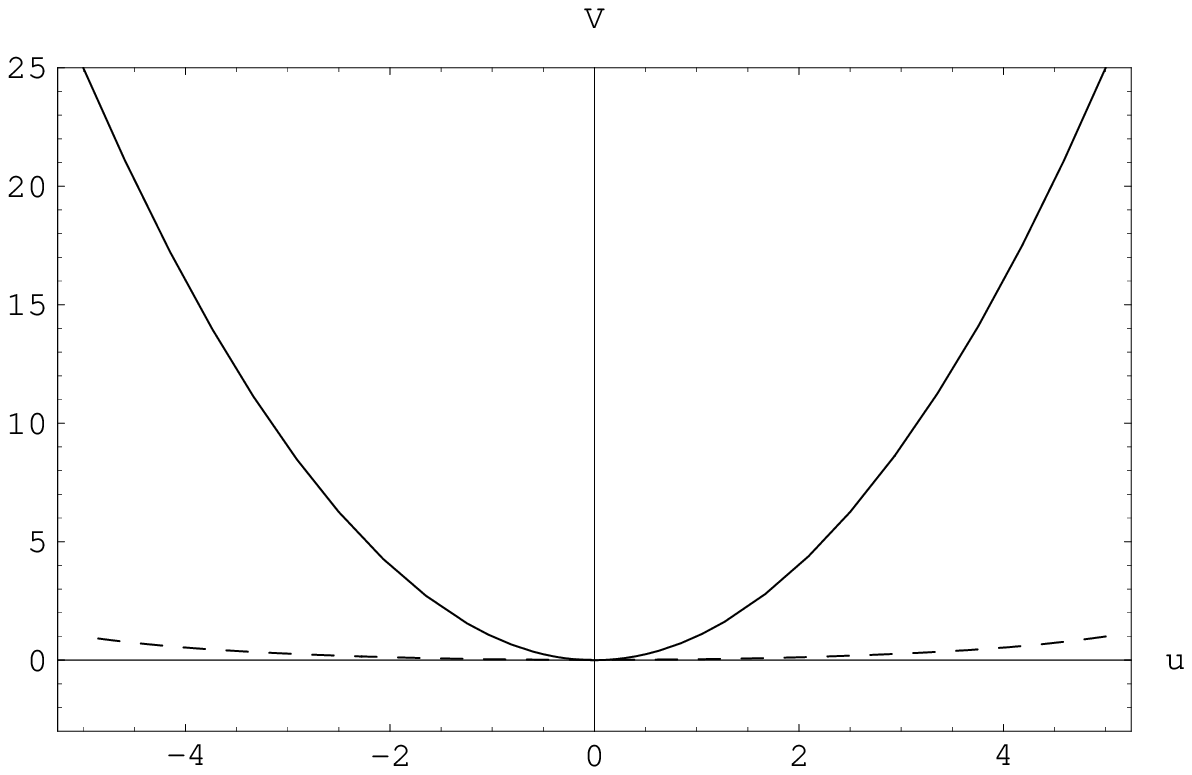}
 &\hspace{2.cm}&
\includegraphics[width=8cm]{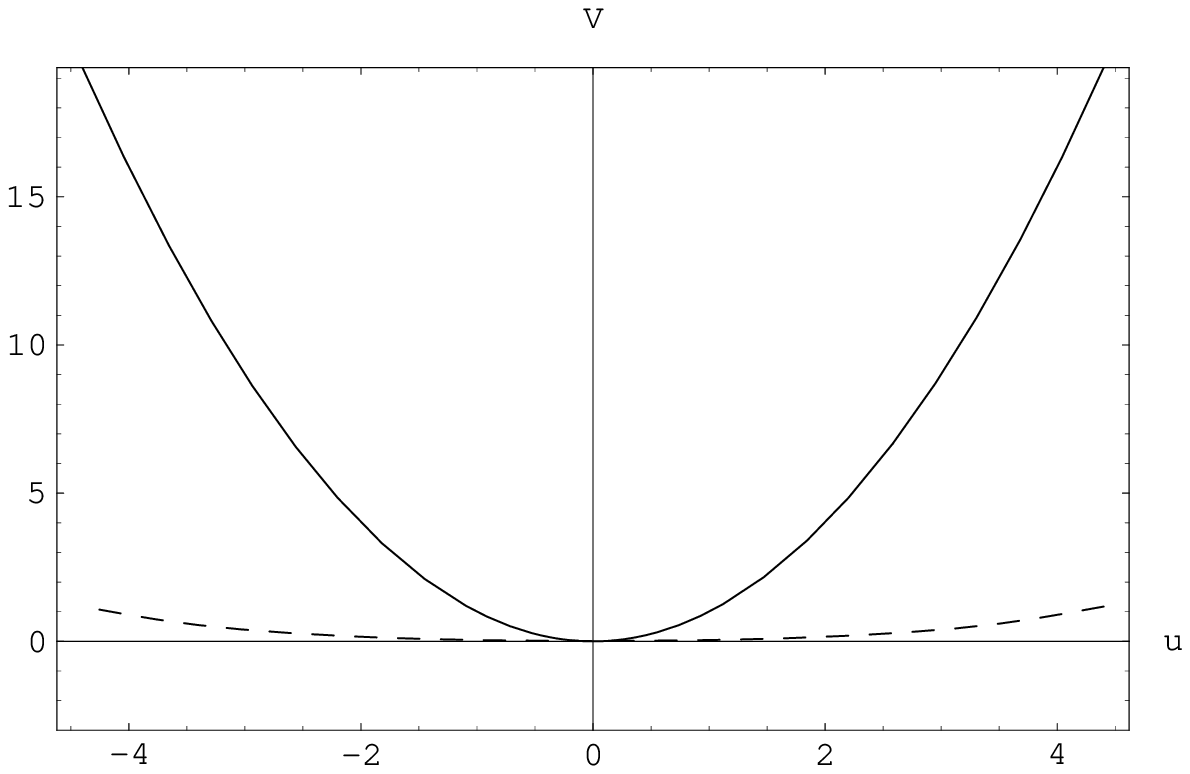}
\end{tabular}}
\caption{$f_1(u,v)$: Classical (solid line) and quantum mechanical
(dashed line) potentials for two types of initial conditions: Left,
$A(n)=\frac{\chi^n}{\,{\sqrt{2^n\,n!}}}e^{-\chi^2/4}$ and $\chi=5$;
Right, $A(n)=\frac{n\,\chi^n}{\,{\sqrt{2^n\,n!}}}e^{-\chi^2/4}$ and
$\chi=4$.} \label{fig3b}
\end{figure}

\begin{figure}
\centering
\includegraphics[width=8cm]{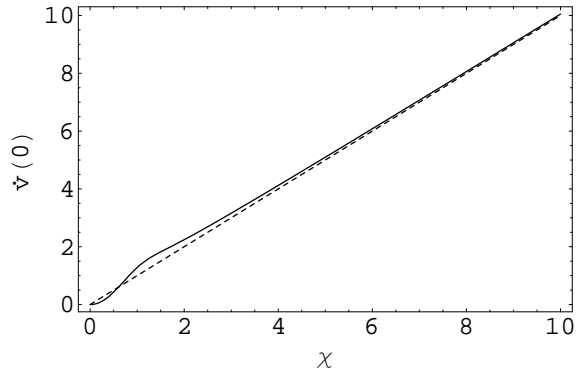}
\caption{$\gamma=2$ case: Classical (dashed line) and Bohmian (solid
line) initial velocity versus $\chi$ for
$A(n)=\frac{\chi^n}{\,{\sqrt{2^n\,n!}}}e^{-\chi^2/4}$.}
\label{fig4b}
\end{figure}

\begin{figure}
\centerline{\begin{tabular}{ccc}
\includegraphics[width=8cm]{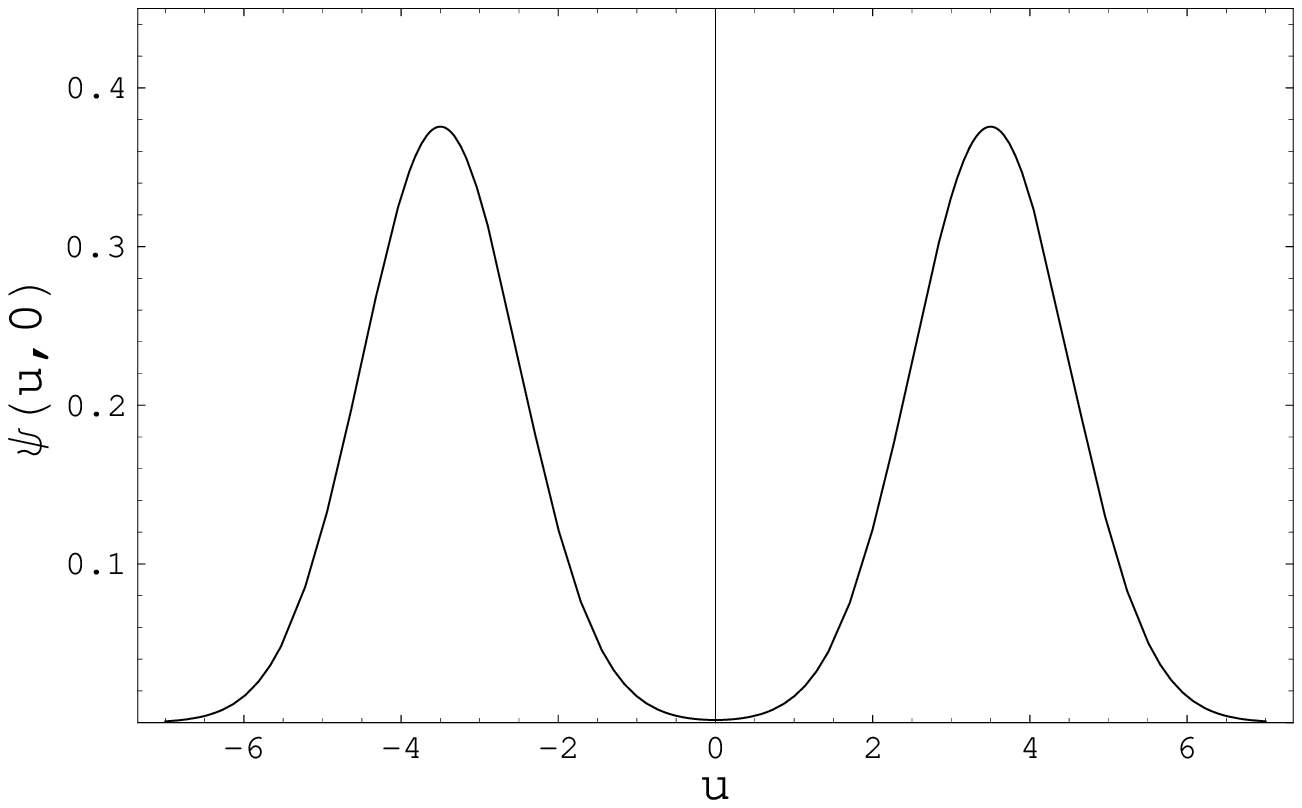}
 &\hspace{2.cm}&
\includegraphics[width=8cm]{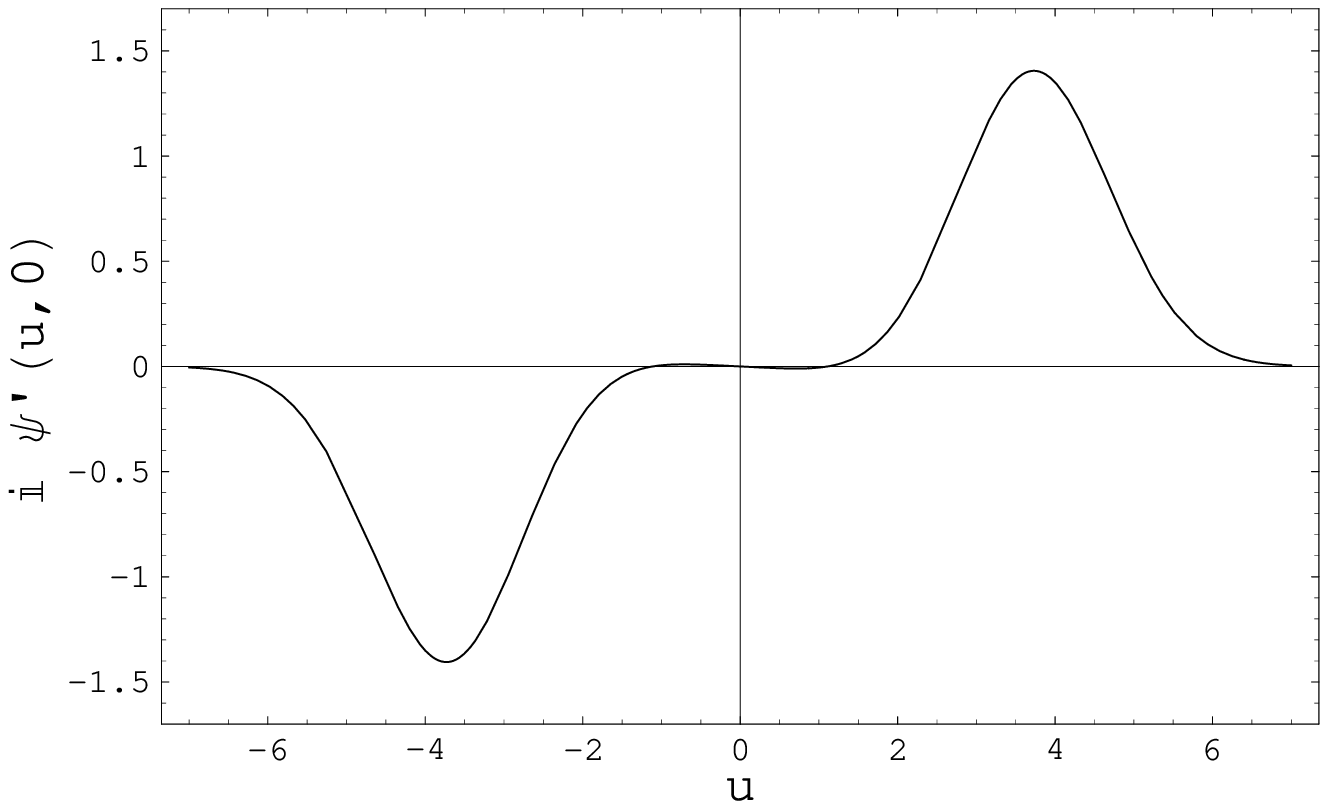}
\end{tabular}}
\caption{$\gamma=2$ case: Left, the initial wave function $
\psi(u,0)$ and right, the initial slope of the wave function $
i\psi'(u,0)$ for
$A(n)=\frac{\,\chi^n}{\,{\sqrt{2^n\,n!}}}e^{-\chi^2/4}$ and
$\chi=3.5$.} \label{figinitial}
\end{figure}

\begin{figure}
\centerline{\begin{tabular}{ccc}
\includegraphics[width=8cm]{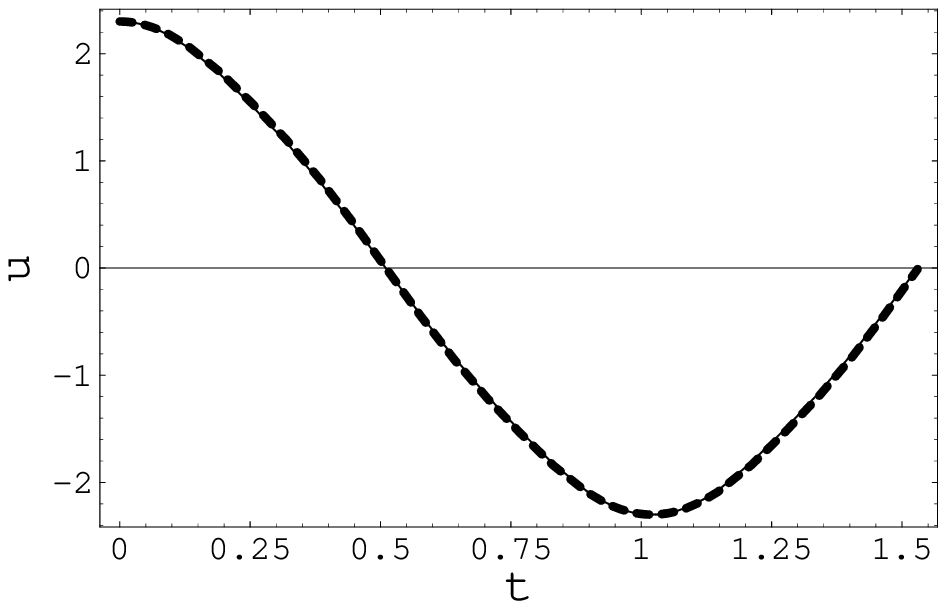}
 &\hspace{2.cm}&
\includegraphics[width=8cm]{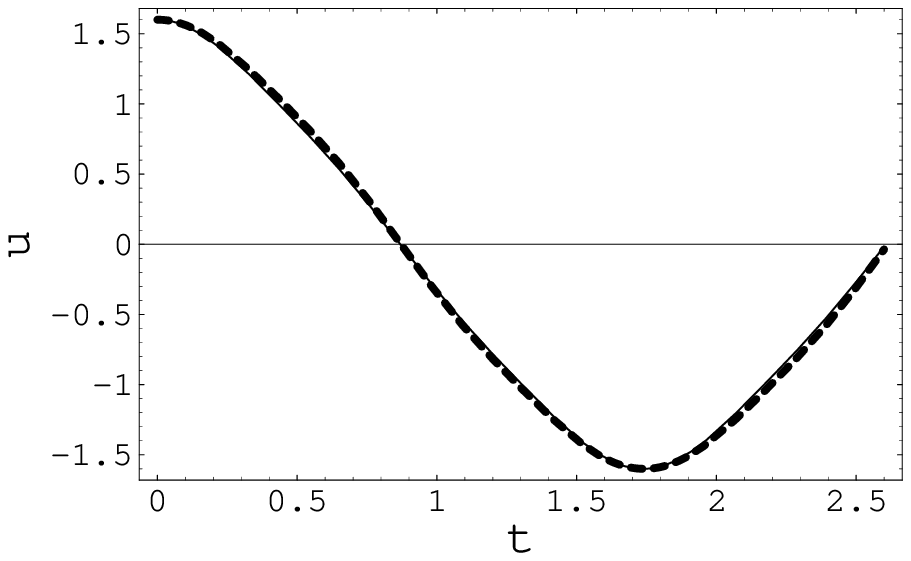}
\end{tabular}}
\caption{Classical and Bohmian values of $u$ versus $t$ for
$\gamma=4$ (left) and $\gamma=6$ (right) with the initial conditions
of Figs. (\ref{fig3},\ref{fig4}), respectively.} \label{figbohm}
\end{figure}

Using the explicit form of the wave packets, these differential
equations can be solved numerically to find the time evolution of
$u$ and $v$. First, consider the case when $\gamma=2$. In this
cases, it is apparent that the full potential for $u$ ($v$) is no
longer equal to $u^2$ ($v^2$) but is
$u^2-\frac{1}{R}\frac{\partial^2 R}{\partial u^2}$
($v^2-\frac{1}{R}\frac{\partial^2 R}{\partial v^2}$). In the right
part of figures \ref{fig1b} and \ref{fig2b}, we have shown the
classical and Bohmian trajectories together for two different
choices of initial wave function
($A(n)=\frac{\,\chi^n}{\,{\sqrt{2^n\,n!}}}e^{-\chi^2/4}$,
$A(n)=\frac{n\,\chi^n}{\,{\sqrt{2^n\,n!}}}e^{-\chi^2/4}$). We see
that the Bohmian trajectories are in good agreement with the
classical counterparts. Now, let us find the quantum potential for
instance in $u$ direction along the Bohmian trajectories which is
given by
\begin{eqnarray}\label{potQ}
V_Q=-\frac{1}{R}\frac{\partial^2 R}{\partial u^2}=-\frac{x'^2+
x\,x'' + {y'}^2 + y\,y'' }{ {x}^2 + {y}^2 }+\left(\frac{ x\,x' +
y\,y' }{ {x}^2 + {y}^2 }\right)^2
\end{eqnarray}
where prime denotes the derivative with respect to $u$. Figure
\ref{fig3b} shows the classical ($V_C$) and quantum ($V_Q$)
potentials for two mentioned initial conditions. In particular, for
$A(n)=\frac{\,\chi^n}{\,{\sqrt{2^n\,n!}}}e^{-\chi^2/4}$, we found
that for $\chi \gtrsim 3$ (where $\chi$ is also the classical radius
of motion for this choice of expansion coefficients) the functional
form of the quantum potential is $V_Q=V_Q(x/\chi)$ with the maximum
value at $x=\chi$. This means that
\begin{eqnarray}\label{VQ/VC}
\frac{V_Q^{\mbox{\footnotesize max}}}{V_C^{\mbox{\footnotesize
max}}}\propto\frac{1}{\chi^2},\hspace{1cm}\mbox{for}\hspace{1cm}\chi
\gtrsim 3.
\end{eqnarray}
Moreover, as indicated in Fig.~\ref{fig4b}, initial Bohmian velocity
coincides well with the classical counterpart for large $\chi$ which
is compatible with the smallness of the quantum potential
(\ref{VQ/VC}). In fact, for this choice of expansion coefficients,
the initial wave function consists of two lumps centered at $u=\pm
\chi$ as (Fig.~\ref{figinitial})
\begin{equation}
\psi(u,0)=\frac{1}{2
\pi^{1/4}}\left(e^{-(u-\chi)^2/2}+e^{-(u+\chi)^2/2}\right).\label{eqpsi0}
\end{equation}
Therefore, the complete classical and quantum correspondence occurs
when there is no significant overlap between the two pieces of
$\psi(u,0)$. This means that to have a good correspondence for small
radii, we need to choose a different set of coefficients or initial
wave function which leads to a more localized wave function with
infinitesimal overlap between its parts. We can also use causal
interpretation for other cases. In particular, Fig.~\ref{figbohm}
shows the obtained Bohmian positions versus time (\textit{i.e.}
$u(t)$) for $\gamma=4$ and $\gamma=6$ which coincide well with their
classical counterparts.

\section{Conclusions}\label{sec4}
We have described a Stephani type cosmology near its symmetry center
leading to classical dynamical equations given by
(\ref{u''}-\ref{u2}) and the corresponding WDW equation represented
by (\ref{eq10}). All these equations are not exactly solvable and we
have solved these equations numerically by an implementation of the
SM for the quantum cosmology cases. We then constructed the wave
packets via canonical proposal which exhibit a good
classical-quantum correspondence. This method proposes a particular
connection between position and momentum distributions which
correspond to their classical quantities and respect to the
uncertainty principle at the same time. Here, using canonical
prescription, we tried to construct the wave packets which peak
around the classical trajectories and simulate their classical
counterparts. We have also studied the situation using de-Broglie
Bohm interpretation of quantum mechanics to quantify our purpose of
classical and quantum correspondence and showed that the Bohmian
positions and momenta coincide well with their classical values upon
choosing arbitrary but appropriate initial conditions. Moreover, We
showed that, in some cases, contrary to FRW cases, the bound state
solutions also exist for all positive values of the cosmological
constant.

\end{document}